\documentclass[10pt,aps,prl,twocolumn,superscriptaddress]{revtex4-2}
\usepackage[x11names]{xcolor}
\usepackage{amsmath,amssymb}
\usepackage{amsthm}
\usepackage{graphicx} 
\usepackage{soul}
\usepackage{enumitem}

\usepackage[colorlinks=true,
            linkcolor=black,
            citecolor=black,
            urlcolor=black,
            bookmarks=true]{hyperref}
\newcommand{\LO}{\mathsf{LO}}
\newcommand{\LOCC}{\mathsf{LOCC}}
\newcommand{\toLOCC}{\underset{\LOCC}{\longrightarrow}}
\newcommand{\toLO}{\underset{\LO}{\longrightarrow}}

\newcommand{\ot}{\otimes}

\theoremstyle{plain}
\newtheorem{theo}{Theorem}
\newtheorem{prop}[theo]{Proposition}
\newtheorem{lem}[theo]{Lemma}

\newtheorem{defi}[theo]{Definition}
\newtheorem{cor}[theo]{Corollary}

\newcommand{\pa}[1]{\left( #1 \right)}
\newcommand{\pac}[1]{\left[ #1 \right]}
\newcommand{\paa}[1]{\left\{ #1 \right\}}

\newcommand{\SpecFun}{\mathsf{SpecFun}}
\renewcommand{\S}{\mathcal{S}}
\renewcommand{\P}{\mathcal{P}}
\newcommand{\States}{\mathcal{M}}

\newcommand{\bra}[1]{\left\langle #1 \right\vert} 			
\newcommand{\ket}[1]{\left\vert #1 \right\rangle} 			

\newcommand\eqdef{\mathrel{\overset{\makebox[0pt]{\mbox{\normalfont\tiny\sffamily def}}}{=}}}
\newcommand{\bb}[1]{\mathbb{#1}}

\newcommand{\GHZ}{\mathrm{GHZ}}
\newcommand{\sGHZ}{\mathrm{sGHZ}}
\renewcommand{\H}{\mathcal{H}}
\newcommand{\LU}{\mathsf{LU}}
\newcommand{\lat}[1]{\textsl{#1}}
\newcommand{\ie}{\lat{i.e.}~}

\renewcommand{\bar}[1]{\overline{#1}}
\newcommand{\simLU}{\underset{\LU}{\sim}}
\newcommand{\abs}[1]{\left\lvert #1 \right\rvert}
\newcommand{\SpreadOne}{\mathsf{P}}
\newcommand{\SpreadTwo}{\mathsf{Q}}

\newcommand{\GM}{\mathsf{GM}}
\newcommand{\eg}{\lat{e.g.}~}
\DeclareMathOperator{\Tr}{Tr}          

\makeatletter
\newcommand{\customlabel}[2]{%
    \phantomsection
    \def\@currentlabel{#1}%
    \label{#2}%
}
\makeatother

\begin{document}

\title{Detection of genuine multipartite entanglement by means of entanglement monotones}

\author{Sylvain Carrozza}
\email{sylvain.carrozza@ube.fr}

\author{Johann Chevrier}
\email{johann.chevrier@ube.fr}

\affiliation{Universit\'{e} Bourgogne Europe, CNRS, IMB,\\ UMR 5584, 21000 Dijon, France}

\author{Luca Lionni}
\email{luca.lionni@ens-lyon.fr}

\affiliation{CNRS, ENS de Lyon, LPENSL,\\ UMR 5672, 69342 Lyon cedex 07, France
}

\date{\today}

\begin{abstract}
    \noindent We look for pure state entanglement monotones that may serve as certificates of a strong form of genuine multipartite entanglement, that we refer to as \emph{irreducible} entanglement: a $D$-partite state (with $D\geq 3$) is \emph{irreducibly $D$-partite entangled} if it cannot be decomposed as a product of states that are at most $(D-1)$-partite entangled. We start by establishing a no-go result proving that no reasonable genuine multipartite entanglement entropy can simultaneously obey $\LOCC$-monotonicity and subadditivity. After relaxing the first condition to $\LO$-monotonicity, we provide a simple method generating genuine multipartite entanglement monotones with the desired properties. Focusing on a subfamily of these monotones, we show that they are able to detect any form of genuine tripartite entanglement in three-qubit systems.
\end{abstract}

\maketitle

\paragraph{Introduction.} Characterizing the entanglement structure of a multipartite quantum pure state $\ket{\psi}\in \H_1 \ot \cdots \ot \H_D$ ($D \geq 3$) is a difficult problem. Indeed, by contrast with the bipartite case ($D=2$), the polynomial invariants \cite{PhysRevA.58.1833, Makhlin2002Nonlocal, AnthonySudbery_2001, Vrana:2011dka, Turner2017, Carrozza:2026qcf} that one must consider to separate distinct entanglement structures proliferate rapidly, both in terms of the local dimensions $\dim(\H_c)$ (which are assumed to be finite in this letter) \cite{Turner2017} and the polynomial order \cite{BenGeloun:2013lim}. In this context, identifying elementary criteria allowing one to certify that a state belongs to a certain non-trivial class of entangled states is theoretically and practically valuable. For instance, determining whether an arbitrary bipartite mixed state $\rho$ (which can be purified to a tripartite pure state) is separable or entangled is known to be an NP-complete problem \cite{10.1145/780542.780545}, but simple certificates of entanglement do exist in this context: if the partial transpose of $\rho$ admits a non-positive eigenvalue, then $\rho$ is necessarily entangled \cite{PhysRevLett.77.1413, HORODECKI19961}, and this can be established by studying a subfamily of polynomial invariants that computes the moments of the partial transpose (see \eg \cite{Dong:2021clv, Elben:2020hpu, Mallick:2026bji} and references therein). 

In this letter, we are interested in devising simple tests certifying that $\ket{\psi}$ admits \emph{genuine} multipartite entanglement. 
In the existing literature \cite{Ma:2011yon,Ma:2023ecg,Xie:2021hsy}, a state $\ket\psi \in \H_1 \otimes \cdots \otimes \H_D \equiv \H$ is said to be \emph{genuinely $D$-partite entangled} if it is \emph{not biseparable} \ie one cannot find $B \subset \{1, \ldots , D\}$, $B\neq \emptyset$, such that $\ket\psi$ is separable with respect to the bipartition $\H \simeq \left( \bigotimes_{c \in B}\H_{c}\right) \otimes \left( \bigotimes_{c \notin B}\H_{c}\right)$ of the original $D$ subsystems. We will be more specifically interested in a considerably strengthened version of this notion: $\ket\psi$ will be said to be \emph{irreducibly $D$-partite entangled} if it cannot be decomposed as a product of states that are at most $(D-1)$-partite (see Def.~\ref{def_strong_genuine}). Clearly, an irreducibly $D$-partite entangled state is also genuinely $D$-partite entangled, but the converse is not true.

\medskip

\paragraph{Objectives and results.} We fix $D \geq 3$ throughout the text and denote by $\States_D$ the set of all finite-dimensional $D$-partite pure states: that is, for any $\ket\psi \in \States_D$, there exist finite-dimensional Hilbert spaces $\{\H_c \}_{1\leq c\leq D}$ such that $\ket\psi \in \H_1 \otimes \cdots \otimes \H_D$ and $\langle \psi \vert \psi \rangle=1$. Our main goal is to identify non-trivial functions $\S: \States_D \to [0,+\infty]$ obeying the following three conditions:
\begin{enumerate}[label=(\roman*)]
    \item \label{it:LO} $\S$ is \emph{non-increasing} under \emph{local operations} ($\LO$);
    \item \label{it:sub} $\S$ is \emph{subadditive}, namely: for any $\ket\psi , \ket\phi \in \States_D$,
    \begin{equation}
    \S(\ket\psi \otimes \ket\phi)    \leq \S(\ket\psi) + \S(\ket\phi)\,;
    \end{equation}
    \item \label{it:bisep} $\S$ \emph{vanishes on biseparable states}.
\end{enumerate}
We have allowed $\S$ to take strictly infinite value because some multipartite generalizations of R\'{e}nyi entanglement entropies, as considered in \eg \cite{Carrozza:2026qcf,Dong:2021clv,Yu:2021uzg,Neven:2021igr,Carrasco:2022rcy,Tarabunga:2025znd}, naturally lead to this behavior (in contrast with the bipartite setting). We also note that: \ref{it:LO} implies that $\S$ is \emph{local unitary} ($\LU$) invariant; while~\ref{it:sub} and \ref{it:bisep} together imply that $\S(\ket\psi)=0$ for any $\ket\psi \in \States_D$ that is \emph{reducibly} (\ie \emph{not} irreducibly) $D$-partite entangled. We will also consider a stronger version of condition \ref{it:LO}, denoted (i')\customlabel{(i')}{it:LOCC}, which requires $\S$ to be non-increasing under \emph{local operations and classical communication} ($\LOCC$). The purpose of this letter is twofold. On the one hand, we show that the simultaneous satisfaction of conditions~\ref{it:LOCC}, \ref{it:sub} and \ref{it:bisep} implies that $\S$ vanishes on any \emph{spectral-hypergraph tensor} (SHT) state \cite{Carrozza:2026jth},  and given the generality of the SHT Ansatz, we interpret this first result as a no-go theorem forbidding reasonable implementations of condition \ref{it:LOCC}. On the other hand, we introduce explicit examples of  functions $\S$ satisfying conditions~\ref{it:LO}, \ref{it:sub} and \ref{it:bisep}, that are non-vanishing on any irreducibly $D$-partite entangled SHT state. Any such $\LO$-monotone can serve as a witness of genuine $D$-partite entanglement beyond SHT states: more precisely, $\S(\ket\psi)>0$ implies that $\ket\psi$ is irreducibly $D$-partite entangled. In the three-qubit system, we prove that a converse also holds, so that $\ket\psi \in \mathbb{C}^2\ot \mathbb{C}^2 \ot \mathbb{C}^2$ is irreducibly (resp. genuinely) $3$-partite entangled if and only if $\S(\ket\psi)>0$. 

\medskip

\paragraph{$\LU$ equivalence and trace-invariants.} Two $D$-partite states $\ket\psi, \ket\phi \in \States_D$ that belong to the same Hilbert space $\H_1 \otimes \cdots \otimes \H_D$ are said to be $\LU$-equivalent -- noted $\ket\psi \simLU \ket\phi$ -- if they can be mapped to one-another by a local unitary transformation \ie a linear map of the form $U_1 \otimes \cdots \otimes U_D$, with $U_c$ a unitary on $\H_c$ for any $c$ \cite{Ziman:2001yzu}. More generally, when $\ket\psi$ and $\ket\phi$ do not belong to the same Hilbert space, they are said to be $\LU$-equivalent if they can be identified up to a local isometry (see Def.~2.2 of \cite{Carrozza:2026qcf}). The $D$-partite state space $\States_D$ can be partitioned into $\LU$-equivalence classes (or $\LU$ entanglement structures), which are known to be separated by the algebra of $\LU$-invariant polynomials (see \eg \cite{meyer2002invariants,Carrozza:2026qcf}). A particularly convenient set of generators, known as \emph{trace-invariants}, plays a central role in our work; we briefly recall their definition (see \cite{Carrozza:2026qcf} and references therein for more detail). Any $\ket{\psi} \in \States_D$ can be decomposed relative to a choice of local basis $\{\ket{i_c}\}_{1 \leq i_c \leq N_c}$ in each subsystem $c$ as:
\begin{equation}
    \ket{\psi} = \sum_{i_1 = 1}^{N_1} \cdots \sum_{i_D = 1}^{N_D} \psi_{i_1 \dots i_D} \ket{i_1} \ot \cdots \otimes \ket{i_D} \,.
\end{equation} Trace-invariants are homogeneous polynomials in the tensor components $\psi_{i_1 \dots i_D}$ and their complex conjugates $\bar \psi_{j_1 \dots j_D}$, and are in one-to-one correspondence with \textit{bipartite regular $D$-edge-colored graphs} or, simply, \emph{colored graphs} (see \eg \cite{Gurau:2011xp, Bonzom:2012hw,gurau_random_2017,Collins2025,Carrozza:2026qcf} and references therein). Each such graph $G$ consists in a set of white vertices $V_\circ$, a set of black vertices $V_\bullet$ and, for any $1\leq c \leq D$, a set of edges of \emph{color} $c$ denoted $E_c$. Furthermore, two combinatorial constraints are imposed: a) any edge $e$ must connect a white vertex $v$ to a black vertex $w$, and can be defined as the vertex pair $e=(v,w)\in V_\circ \times V_\bullet$; b) to any vertex $v$ is incident a unique edge of a given color $c$ \ie in the set of vertex pairs $E_c$, $v$ appears exactly once. In particular, the sets $V_\circ$, $V_\bullet$ and $E_c$ (for any $c$) must have the same cardinality, which we denote by $k(G)\in \mathbb{N}^*$. A graph $G$ is associated with a unique homogeneous polynomial of degree $k(G)$, or trace-invariant, denoted $\Tr_G$. The evaluation $\Tr_G (\ket\psi)$ can be obtained as a tensor network contraction of $k(G)$ copies of the tensor components of $\ket\psi$ with $k(G)$ copies of their conjugates,  as follows: 1) each white (resp.~black) vertex $v$ (resp.~$w$) is associated with a factor $\psi_{i_1^{(v)} \dots i_D^{(v)}}$ (resp.~$\bar \psi_{j_1^{(w)} \dots j_D^{(w)}}$); 2) each color-$c$ edge $e_c = (v,w) \in E_c$ is associated with a Kronecker delta $\delta_{i_c^{(v)}, j_c^{(w)}}$; 3) all the indices are then summed over their range. For example, the norm squared in the tripartite case (\ie $D=3$), is given by:
\begin{equation} \label{eq:melon}
    \Tr_{G}(\ket{\psi}) = \sum_{{\color{Red4}i_1} = 1}^{N_1} \sum_{{\color{Blue4}i_2} = 1}^{N_2} \sum_{{\color{Green4}i_3} = 1}^{N_3} \psi_{{\color{Red4}i_1} {\color{Blue4}i_2} {\color{Green4}i_3}} \bar{\psi}_{{\color{Red4}i_1} {\color{Blue4}i_2} {\color{Green4}i_3}} = \langle \psi \vert \psi \rangle = 1\,,
\end{equation}
where $G = \vcenter{\hbox{\includegraphics[scale = 1]{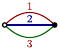}}}$.

Local operations, introduced in condition~\ref{it:LO}, can be characterized in terms of trace-invariants \cite{Carrozza:2026qcf}. More precisely, for any $\ket\psi , \ket\varphi \in \States_D$, we have $\ket \psi \toLO \ket \varphi$ if and only if there exists a pure state $\ket \eta \in \States_D$ such that, for any colored graph $G$: $\Tr_G(\ket \psi) = \Tr_G(\ket \eta) \Tr_G(\ket \varphi)$.

\medskip

\paragraph{Spectral hypergraph-tensor states.} Denoting by  $\P_D \eqdef \paa{B \subset \paa{1,\dots,D} \, \vert \, \abs{B} \geq 2}$ and $1_D \eqdef \paa{1,\dots,D}$, let us recall the definition of two families of ($\LU$-equivalence classes of) states considered in \cite{Carrozza:2026qcf,Carrozza:2026jth}. For any $B \in \P_D$ and $N \in \mathbb{N}^*$ , we denote by $\ket{\GHZ}_{B, N}$ a $\GHZ$ state of dimension $N$ with support on $B$ \ie in suitable local bases, we have
\begin{equation}
    \ket{\GHZ}_{B,N} \eqdef \frac{1}{\sqrt{N}} \sum_{i = 1}^{N} \ket{i}^{\ot \abs{B}} \,.
\end{equation}
Next, given a \textit{weight function} $\alpha: B \mapsto \alpha(B)$, that associates a dimension parameter $\alpha(B) \in \bb{N}^*$ with each element $B \in \P_D$, the \emph{hypergraph-tensor} (HT) state $\ket{\psi_\alpha}$ is defined as
\begin{equation}
    \ket{\psi_\alpha} \eqdef \bigotimes_{B \in \P_D} \ket{\GHZ}_{B,\alpha(B)} \,.
\end{equation}
The set of weight functions is denoted by $W_D$, each element of which labels a distinct $\LU$-orbit in $\States_D$ \cite{Carrozza:2026qcf}. More broadly, with any $B \subset \P_D$, and any decreasing sequence $\pmb{p}= (p_i)_{i \in \mathbb{N}^*}$ of non-negative reals, with finitely many non-zero entries that sum to $1$, we associate a \emph{spectral} $\GHZ$ state of the form:
\begin{equation}
    \ket{\sGHZ}_{B,\, \pmb{p}} \eqdef \sum_{i \in \bb{N}^*} \sqrt{p_i} \ket{i}^{\ot \abs{B}} \,.
\end{equation}
$\pmb{p}$ is referred to as the \emph{entanglement spectrum} of this state, and the number of non-zero entries in $\pmb{p}$ as its \emph{rank}. Given a \textit{spectral function} $\pmb{p}: B \mapsto \pmb{p}(B)$ that attributes a spectrum to each element $B \in \P_D$, the \textit{spectral hypergraph-tensor} (SHT) state $\ket{\psi_{\pmb{p}}}$ is defined as:
\begin{equation}\label{eq:SHT}
    \ket{\psi_{\pmb{p}}} \eqdef \bigotimes_{B \in \P_D} \ket{\sGHZ}_{B, \, \pmb{p}(B)} \,.
\end{equation}
The set of spectral functions is denoted by $\SpecFun_D$, and labels inequivalent $\LU$-orbits in $\States_D$ \cite{Carrozza:2026jth}. Given that, for any $N\in \mathbb{N}^*$ and $B\in\P_D$, 
\begin{equation}
    \ket\sGHZ_{B, \left(1/N, \ldots , 1/N, 0, \ldots\right)} = \ket\GHZ_{B,N}\,,
\end{equation}
HT states are nothing but SHT states with a flat entanglement spectrum on every block $B$ \cite{Carrozza:2026jth}. 

\medskip

\paragraph{Irreducible $D$-partite entanglement.} We are ready to formally introduce this notion.
\begin{defi}\label{def_strong_genuine}
    A $D$-partite state $\ket\psi \in \States_D$ is said to be \emph{irreducibly $D$-partite entangled} if one cannot find states $\{\ket{\eta_{B}}\}_{B \subsetneq \{1, \ldots , D\}}$ such that 
    \begin{equation}
        \ket\psi \simLU \bigotimes_{B \subsetneq \{1, \ldots , D\}} \ket{\eta_B} 
    \end{equation}
    and, for any $B$, $\ket{\eta_B}$ is supported on $B$.
\end{defi}
\noindent Note that \emph{irreducible} $D$-partite entanglement implies \emph{genuine} $D$-partite entanglement. On the other hand, for $N\geq 2$, the \emph{triangle state}
\begin{equation}\label{eq:triangle_state}
\ket\GHZ_{\{1,2\}, N} \otimes \ket\GHZ_{\{1,3\}, N} \otimes \ket\GHZ_{\{2,3\}, N}     
\end{equation}
is genuinely $3$-partite entangled but reducibly $3$-partite entangled, showing that the converse does not hold.

\medskip

\paragraph{Quantum teleportation protocol.} In Prop.~5.13 of \cite{Carrozza:2026qcf}, it was shown that, for any $N \in \bb{N}^*$, any $B_1,B_2, B \in \P_D$ such that $B_1 \cap B_2 \neq \emptyset$ and $B = B_1 \cup B_2$, the state $\ket{\GHZ}_{B_1,N} \ot \ket{\GHZ}_{B_2,N}$ can be turned into $\ket{\GHZ}_{B,N}$ by means of a $\LOCC$ transformation, which we denote
\begin{equation} \label{eq:LOCC_QT}
    \ket{\GHZ}_{B_1,N} \ot \ket{\GHZ}_{B_2,N} \toLOCC \ket{\GHZ}_{B,N}\,.
\end{equation}
This was implemented by a slight adaptation of the quantum teleportation protocol \cite{PhysRevLett.70.1895}, which we generalize further to incorporate SHT states. \begin{lem} \label{lem:LOCC}
    Let $N \in \bb{N}^*$, $B_1,B_2 \in \P_D$ with $B_1 \cap B_2 \neq \emptyset$ and $B = B_1 \cup B_2$. For any \emph{entanglement spectrum} $\pmb{p}$ of rank at most $N$, we have:
    \begin{equation}\label{eq:LOCC-sGHZ}
        \ket{\GHZ}_{B_1,N} \ot \ket{\GHZ}_{B_2,N} \toLOCC \ket{\sGHZ}_{B,\,\pmb{p}}\,.
    \end{equation}
\end{lem}

\begin{proof}
    See Appendix A.
\end{proof}

\paragraph{No--go theorem on $\LOCC$-monotonicity.} A \emph{Bell state} is a state of the form $\ket{\GHZ}_{B,N}$ with $\abs{B}=2$ and $N \in \mathbb{N}^*$. Since $D\geq 3$, any such state is \textit{biseparable} (it is even completely separable when $N=1$).
Our first theorem demonstrates that requiring a multipartite entanglement entropy $\S$ to vanish on Bell states (a weakened version of condition~\ref{it:bisep}), while obeying \ref{it:LOCC} and \ref{it:sub}, has so stringent consequences that they make $\S$ largely ineffective as a witness of irreducible $D$-partite entanglement.
\begin{theo} \label{thm_1}
    Let $\S: \States_D \to [0, +\infty]$ a $\LU$-invariant function. If $\S$ is non-increasing under $\LOCC$ transformations, is subadditive, and vanishes on all Bell states, then:
    \begin{equation}
        \forall \pmb{p} \in \SpecFun_D \,, \qquad \S(\ket{\psi_{\pmb{p}}}) = 0 \,.
    \end{equation}
This conclusion holds in particular whenever conditions~\ref{it:LOCC}, \ref{it:sub} and \ref{it:bisep} above are assumed.    
\end{theo}

\begin{proof}
    Let $N\in \bb{N}^*$ and let $\ket{\Phi_c}$ denote the $c$-Star state introduced in \cite[p.~53]{Carrozza:2026qcf}, defined as:
    \begin{equation}
        \ket{\Phi_c} \eqdef \bigotimes_{c' \in \{1,\ldots , D\} \setminus\{c\}} \ket{\GHZ}_{\{c,c'\},N} \,.
    \end{equation}
    The hypothesis that $\S$ vanishes for all Bell states combined with the subadditivity assumption, yields:
    \begin{equation}\label{eq:thm1-1}
        0 = \sum_{c' \neq c} \S(\ket{\GHZ}_{\{c,c'\},N}) \geq \S(\ket{\Phi_c})  \,.
    \end{equation}
    Furthermore, by repeated use of the $\LOCC$ relation \eqref{eq:LOCC_QT}, we can infer that $\ket{\Phi_c} \toLOCC \ket{\GHZ}_{1_D, N}$. 
    Thus, the monotonicity and non-negativity assumptions lead to $\S(\ket{\Phi_c}) \geq \S(\ket{\GHZ}_{1_D,N}) \geq 0$. Combined with \eqref{eq:thm1-1}, this imposes $\S(\ket{\GHZ}_{1_D,N}) = 0$.\\
    Next, by Prop.~5.11 of \cite{Carrozza:2026qcf}, we know that $\ket{\GHZ}_{1_D,N}  \toLOCC \ket{\GHZ}_{B,N}$ for any $B \in \P_D\setminus \{1_D\}$ and any $N \in \bb{N}^*$; thus, by monotonicity, we must have $\S(\ket{\GHZ}_{B,N}) = 0$. \\
    Moreover, for any block $B \in \P_D$ and any spectrum $\pmb{p}$ of rank $N\in \mathbb{N}^*$, Lem.~\ref{lem:LOCC} implies the $\LOCC$ relation \eqref{eq:LOCC-sGHZ} specialized to $B_1=B=B_2$. Subadditivity, combined with non-negativity, thus leads to
    \begin{equation}
        0 = 2\S(\ket{\GHZ}_{B,N})\geq \S(\ket{\GHZ}_{B,\pmb{p}}) \geq 0\,,
    \end{equation}
    and therefore $\S(\ket{\GHZ}_{B,\pmb{p}})=0$. \\
    We can now prove that $\S$ vanishes for all SHT states. Let $\pmb{p} \in \SpecFun_D$. Applying $\S$ to equation \eqref{eq:SHT}, while invoking subadditivity and non-negativity, one obtains
    \begin{equation}
        0 = \sum_{B \in \P_D} \S(\ket{\GHZ}_{B,\pmb{p}(B)}) \geq \S(\ket{\psi_{\pmb{p}}}) \geq 0 \,,
    \end{equation}
    hence $\S(\ket{\psi_{\pmb{p}}}) =0$.
\end{proof}

\noindent Condition~\ref{it:bisep} (or its weaker version, requiring $\S$ to vanish on all Bell states) is often imposed in the quantum information and holography literature: it is indeed the defining property of any \textit{genuine multipartite entanglement entropy} (GME) \cite{Ma:2011yon,Ma:2023ecg,Xie:2021hsy}. Since it vanishes on any biseparable state, a GME can be used to certify genuine $D$-partite entanglement. Moreover, since the $\GHZ$ state $\ket{\GHZ}_{1_D, N}$ (with $N\geq 2$) is the archetypical example of genuinely $D$-partite entangled state, it is natural to require any GME $\S$ to detect the genuine $D$-partite entanglement of this state \ie $\S(\ket{\GHZ}_{1_D, N})>0$. Thm.~\ref{thm_1} proves that this cannot be achieved while preserving conditions \ref{it:LOCC}, \ref{it:sub} and \ref{it:bisep}. 

For example, while GMEs based on concurrences \cite{Ma:2011yon,PhysRevResearch.4.023059,e26070535,Shi:2022itc}, three-qubit teleportation capability \cite{Choi:2022lge} or bipartite entropy \cite{PhysRevA.67.052107}, all detect the genuine entanglement of $\ket{\GHZ}_{1_D, N}$, they fail to be subadditive.
Likewise, the \emph{generalized geometric measure of entanglement} \cite{SenDe:2009ofr,Das:2015wyv}, defined via a maximization of the overlap with the set of biseparable states \footnote{Similarly to how the \emph{geometric measure of entanglement} \cite{Shimony:1995xkl, Linden:2001fkw, Wei:2003qfk}, also known as the \emph{injective norm} (see \eg \cite{Dartois:2024zuc}), is defined via a maximization over completely separable states.}, is known to define a $\LOCC$-monotone but, in compliance with Thm.~\ref{thm_1}, it is not subadditive: indeed, it is non-vanishing on any state that fails to be biseparable (this includes states that have genuine but reducible $D$-partite entanglement, like \eqref{eq:triangle_state}). 

On the other hand, consider the examples of \textit{multipartite entanglement spreads} introduced in \cite{Carrozza:2026jth} and denoted there as $\SpreadOne_n^{(c)}$, $\SpreadTwo_{m,n,p,q}^{(c)}$ and $\mathsf{D}_G^{(\mathrm{MST})}$  (see \cite{Carrozza:2026jth} for proper definitions). Those functions from $\States_D$ to $\pac{0,+\infty}$ are additive (hence, subadditive), non-increasing under $\LO$-transformations, and vanish on all HT states (thus on all Bell states), but do not vanish on all SHT states. By contraposition of Thm.~\ref{thm_1}, we conclude that they all fail to define $\LOCC$-monotones.

A simple corollary of Thm.~\ref{thm_1} allows one to prove that other GMEs considered in the literature cannot define $\LOCC$-monotones.
\begin{cor} \label{thm_2}
    Let $\S: \States_D \to [-\infty , + \infty]$ a $\LU$-invariant function. If $\S$ is additive, vanishes on all Bell states (including separable states), and defines a $\LOCC$-monotone, then: 
    \begin{equation}
        \forall \pmb{p} \in \SpecFun_D \,, \qquad \S(\ket{\psi_{\pmb{p}}}) = 0 \,.
    \end{equation}
\end{cor}

\begin{proof}
    Let us first assume that $\S$ is non-increasing under $\LOCC$. Since any state $\ket\psi \in \States_D$ can be converted into some separable state $\ket{\mathrm{Sep}}$ by $\LOCC$, we have
    \begin{equation}
        \S(\ket{\psi}) \geq \S(\ket{\mathrm{Sep}}) =0\,.
    \end{equation}
    Thus, Thm.~\ref{thm_1} applies. If $\S$ is non-decreasing under $\LOCC$, we can instead apply Thm.~\ref{thm_1} to $- \S$, and reach the same conclusion.
\end{proof}

A direct application of the last theorem concerns the $D$-partite \textit{genuine multi-entropies} introduced in~\cite{Gadde:2022cqi}. They are obtained as an analytic continuation of additive GME Rényi entropies denoted $\GM_{n}^{(D)}$ and called \textit{genuine Rényi multi-entropies} (see \eg~\cite{Iizuka:2025ioc, Iizuka:2025caq}). The latter are obtained by linear combinations of logarithms of trace-invariants. For instance, $\GM_{2}^{(3)}$ is given by the following linear combination:
\begin{equation*}
    \GM_2^{(3)}(\ket{\psi}) = \frac{1}{2}\ln \Tr_{\vcenter{\hbox{\includegraphics[scale = .6]{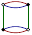}}}\vcenter{\hbox{\includegraphics[scale = .6]{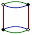}}}\vcenter{\hbox{\includegraphics[scale = .6]{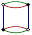}}}}(\ket{\psi}) - \frac{1}{2} \ln  \Tr_{\vcenter{\hbox{\includegraphics[scale = .6]{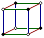}}}} (\ket{\psi}) \,,
\end{equation*}
where we relied on the colored graph  encoding of trace-invariants.

Let $n\geq 2$. Introducing the one-parameter family of entanglement spectra $\pmb{p}_q = (1-q,q,0,\ldots)$, with $q \in \pac{0,1}$, a computation yields:
\begin{equation} \label{eq:GM_n^3}
    \GM_{n}^{(3)}(\ket{\sGHZ}_{1_3,\,\pmb{p}_q}) \underset{q \to 0}{=} \frac{nq}{2(1-n)} + o(q) \,.
\end{equation}
Hence, $\GM_n^{(3)}$ does not vanish on all SHT states, and therefore, by Cor.~\ref{thm_2}, it does not define a $\LOCC$-monotone.  

For $D=4$, $\GM_n^{(4)}$ has a free parameter $a_n$ (see Eq.~(2.9) of~\cite{Iizuka:2025caq}). However, by Cor.~\ref{thm_2}, requiring $\GM_n^{(4)}(\ket{\psi_\alpha}) = 0$ for any $\alpha \in W_D$ is necessary for $\GM_n^{(4)}$ to be $\LOCC$-monotone. For $n>1$, this condition imposes that:
\begin{equation}\label{eq:a_n}
    a_n =  \frac 1 3 - \frac 1 n + \frac{1}{n^2}\,.
\end{equation}
On the other hand, one can show that $\GM_{n}^{(4)}(\ket{\sGHZ}_{1_4,\,\pmb{p}_q}) \underset{q \to 0}{=} \frac{nq}{1-n}(a_n - \frac{1}{3}) + o(q)$, and thus, Cor.~\ref{thm_2} imposes that $a_n = \frac{1}{3}$. This is in contradiction with \eqref{eq:a_n}, hence there is no choice of the free parameter $a_n$ making $\GM_n^{(4)}$ a $\LOCC$-monotone.

For $D = 5$, the same contradiction holds. Indeed, fixing an integer $n>1$, $\GM_n^{(5)}$ has again a free parameter $b_n$ (denoted as $c_5$ in~\cite{Iizuka:2025caq}, Eq.~(3.50)). Firstly, the necessary condition on HT states imposes that
\begin{equation}
    b_n = \frac{1-n}{2n^2} \neq 0\,.
\end{equation}
Secondly, a short computation leads to
\begin{equation}
    \GM_n^{(5)}(\ket{\sGHZ}_{1_5,\,\pmb{p}_q}) \underset{q \to 0}{=} - \frac{nq}{1-q} b_n + o(q) \,,
\end{equation}
which does not vanish identically. It follows from Cor.~\ref{thm_2} that $\GM_n^{(5)}$ is not a $\LOCC$-monotone.

Although we have not checked this in full, we expect that $\GM_n^{(D)}$ fails to be $\LOCC$-monotone for any choice of integers $n\geq 2$ and $D \geq 3$. We also note that our proofs leave the possibility open for these quantities to define $\LOCC$-monotones in a suitable $n\to 1$ limit, an interesting possibility we leave open for future work. However, see \cite{Akella:2026nbc}, where a prescription for the $n\to 1$ limit is provided, relative to which R\'{e}nyi multi-entropies actually \emph{lose} some of the $\LO$-monotonicity properties they admit at any integer value $n\geq 2$.

\medskip

\paragraph{GME entropies and $\LO$-monotonicity.} From the last paragraph, we conclude that any GME detecting the $D$-partite entanglement of a complete $\GHZ$ state is not $\LOCC$-monotone. The next best thing we can do is to require $\LO$-monotonicity, and the purpose of this paragraph is to provide simple examples of entropies that do indeed obey conditions~\ref{it:LO}, \ref{it:sub} and \ref{it:bisep}. 

Let $\sigma \in S_D$, where $S_D$ is the symmetric group on $\{1, \ldots, D\}$. We define $M_\sigma$ and $\tilde{X}_\sigma$ to be the open colored graphs:
\begin{align}
    M_\sigma &\eqdef \vcenter{\hbox{\includegraphics[height = 1.5cm]{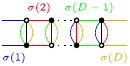}}}\,, \\
    \tilde{X}_\sigma &\eqdef \vcenter{\hbox{\includegraphics[height = 1.5cm]{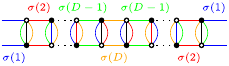}}}\,.
\end{align}
Generalizing the correspondence between colored graphs and polynomials to open colored graphs (in the obvious way, see \cite{Carrozza:2026jth}), given $\ket\psi \in \H_1\otimes \cdots \otimes \H_D$ we can interpret $\Tr_{M_\sigma}(\ket\psi)$ as an endomorphism $A_\psi:\H_{\sigma(1)}\otimes \H_{\sigma(1)}^*\to\H_{\sigma(D)}^*\otimes \H_{\sigma(D)}$, and $\Tr_{\tilde{X}_\sigma}(\ket\psi)$ as $A_{\psi}(A_{\psi})^\dagger$. For any $n \in \bb{N}^*$, we denote by $G_n(\tilde X_\sigma)$ the $D$-colored graph obtained by cyclically gluing $n$ copies of $\tilde X_\sigma$ as illustrated here:
\begin{equation}
     G_n(\tilde X_\sigma) = \vcenter{\hbox{\includegraphics[scale = 1.2]{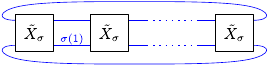}}}\,,
\end{equation}
so that $\Tr_{G_n(\tilde X_\sigma)}(\ket\psi)= \Tr\left( \left(A_\psi (A_\psi)^\dagger \right)^n\right)$. Note that the degree of this invariant is $k(G_n(\tilde X_\sigma))= 2(D-1)n$.
\begin{lem}\label{lem:order_rel}
    For any $n \geq 2$ and $\ket\psi \in \States_D$, we have:
    \begin{equation}
        0 < \left( \Tr_{G_1(\tilde X_\sigma)}(\ket\psi) \right)^n \leq \Tr_{G_n(\tilde X_\sigma)}(\ket\psi) \,.
    \end{equation}
\end{lem} 
\begin{proof}
    Denoting by $\lambda_1, \ldots , \lambda_r \in \mathbb{R}_+^*$ the non-zero eigenvalues of the positive operator $A_{\psi}(A_{\psi})^\dagger$, we have
    \begin{align}
        \left( \Tr_{G_1(\tilde X_\sigma)}(\ket\psi) \right)^n = \left( \sum_{i=1}^r \lambda_i \right)^n &\leq \sum_{i=1}^r (\lambda_i)^n \\ &= \Tr_{G_n(\tilde X_\sigma)}(\ket\psi) \,. \nonumber
    \end{align}
    Moreover, viewing $\Tr_{G_1(\tilde X_\sigma)}(\ket\psi)$ as an inner product, we can derive a Cauchy-Schwarz inequality of the form
    \begin{equation}\label{eq:power_law_bound}
       \abs{\Tr_C (\ket\psi)}^2 \leq \Tr_{G_1(\tilde X_\sigma)}(\ket\psi) \times M\,,
    \end{equation}
    with $M$ a non-zero integer (which depends on the local dimensions $\dim(\H_c)$) and $C$ a so-called \emph{cyclic} graph \cite{Carrozza:2026qcf}. It was shown in Prop.~6.16 of \cite{Carrozza:2026qcf} that cyclic graphs define strictly positive invariants, hence so does $G_1(\tilde X_\sigma)$.   
\end{proof}

In view of the previous lemma, we introduce the map $R_\sigma^{(n)}:\States_D \to [0, +\infty[$, defined as: 
\begin{equation}
    R_\sigma^{(n)}(\ket{\psi}) \eqdef - \ln \pac{ \frac{\Tr_{G_n(\tilde X_\sigma)}(\ket{\psi})}{\pa{\Tr_{G_1(\tilde X_\sigma)}(\ket{\psi})}^n}} \,,
\end{equation}
for any $\ket\psi \in \States_D$. We  also introduce a color-symmetric version of this quantity, namely: 
\begin{equation}
        R^{(n)}\eqdef \frac{1}{D!} \sum_{\sigma \in S_D} R_\sigma^{(n)}  \,.
\end{equation}
Let us show that $R_\sigma^{(n)}$ (and thus $R^{(n)}$) is a GME that is, in particular, $\LO$-monotone.
\begin{theo} \label{thm_GME}
    Let $\sigma \in S_D$ and $n \geq 2$. The map $R_\sigma^{(n)}:\States_D \to [0, +\infty[$ obeys conditions~\ref{it:LO}, \ref{it:sub} and \ref{it:bisep}. Furthermore, for any irreducibly $D$-partite entangled SHT state $\ket\psi$, one has $R_\sigma^{(n)}(\ket\psi)>0$.
\end{theo}
\begin{proof}
    Lem.~\ref{lem:order_rel} guarantees that $R_\sigma^{(n)}$ defines an \emph{entanglement spread} in the sense of \cite{Carrozza:2026jth}; in particular, such a quantity obeys condition~\ref{it:LO} by virtue of Prop.~2.11 of \cite{Carrozza:2026jth}. As a linear combination of logarithms of trace-invariants, $R_\sigma^{(n)}$ is also additive, and therefore obeys condition \ref{it:sub}.
    
    Let $B \subsetneq \paa{1,\dots,D}$, and consider a state $\ket{\psi} \in \States_D$ that is biseparable relative to the bipartition $B \sqcup \bar B$, \ie there exist $\ket{\varphi_B}$ supported on $B$ and $\ket{\varphi_{\bar{B}}}$ supported on $\bar B$ such that $\ket{\psi} \simLU \ket{\varphi_B} \ot \ket{\varphi_{\bar{B}}}$. Then, for any colored graph $G$, one finds that
    \begin{equation}
        \Tr_{G}(\ket\psi) = \Tr_{G\vert_B}(\ket{\varphi_B}) \Tr_{G\vert_{\bar B}}(\ket{\varphi_{\bar B}})\,,
    \end{equation}
    where $G\vert_B$ (resp.~$G\vert_{\bar B}$) denotes the graph obtained by removing from $G$ any edge whose color is in $\bar B$ (resp.~in $B$). But $\tilde{X}_\sigma$ was specifically designed to obey the following combinatorial constraints: for any $C\subsetneq \paa{1,\dots,D}$,
    \begin{equation}\label{eq:comb_constraint}
        G_n(\tilde X_\sigma)|_C \simeq \underbrace{G_1(\tilde X_\sigma)|_C \sqcup \cdots \sqcup G_1(\tilde X_\sigma)|_C}_{n\;\mathrm{copies}}\,,
    \end{equation}
    where $\simeq$ denotes a colored graph isomorphism and $\sqcup$ the disjoint union of graphs. Henceforth, it is clear that $R_\sigma^{(n)}(\ket\psi) = 0$ \ie $R_\sigma^{(n)}$ obeys condition~\ref{it:bisep}.

    Finally, an irreducibly $D$-partite entangled SHT state $\ket\psi$ is associated with a spectral function $\pmb{p}$ attributing a non-trivial spectrum $\pmb{p}(1_D) \neq (1,0,\ldots)$ to the block $1_D$. Then, a direct computation shows that 
    \begin{equation}
        R_\sigma^{(n)}(\ket\psi) = - \ln\pac{\frac{\displaystyle\sum_{i=1}^{+\infty} (\pmb{p}(1_D)_i)^{2(D-1)n}}{\pa{\displaystyle\sum_{i=1}^{+\infty} (\pmb{p}(1_D)_i)^{2(D-1)}}^n }}>0
    \end{equation}
\end{proof}
As a direct corollary, $R_\sigma^{(n)}$ may be used to certify irreducible $D$-partite entanglement.
\begin{cor} \label{cor:if_GME}
     Let $\sigma \in S_D$, $n \geq 2$ and $\ket\psi \in \States_D$. If $R_\sigma^{(n)}(\ket\psi)> 0$, then $\ket\psi$ is irreducibly $D$-partite entangled.
\end{cor}
\noindent In general, whether the converse implication also holds (for $R_\sigma^{(n)}$ or its color-symmetric version $R^{(n)}$) remains an open question. However, we were able to explicitly check that it does in the special case of the three-qubit system.
\begin{prop} \label{prop:iff_3qubits}
    Let $\sigma\in S_3$ and $n \geq 2$. For any $\ket\psi \in \mathbb{C}^2 \otimes \mathbb{C}^2 \otimes \mathbb{C}^2$, $\ket\psi$ is irreducibly $D$-partite entangled if and only if $R_\sigma^{(n)}(\ket\psi) > 0$.
\end{prop}
\begin{proof}
    See Appendix B.
\end{proof} 
\noindent In particular, $R_\sigma^{(n)}$ detects the irreducible $D$-partite entanglement of the W state $\ket{W}\propto \ket{001}+\ket{010}+\ket{100}$ \footnote{We further remark that $R_\sigma^{(2)}$ is \textit{proper} \cite{Xie:2021hsy}, since $R_\sigma^{(2)}(\ket\GHZ) = \ln 2 > 4 \ln 3 - 2 \ln 7 = R_\sigma^{(2)}(\ket W)$.}.

The next result shows that, while the GMEs $\{R_\sigma^{(n)}\}$ provide tests of irreducible $D$-partite entanglement that are in principle applicable in any dimension, those tests will become less and less effective on average in the limit of large local dimensions. Indeed, choosing a state $\ket{\varphi_N}$ uniformly at random in a $D$-partite state space of local dimension $N$, we find that, even though $R_\sigma^{(n)}\left(\ket{\varphi_N}\right)>0$ with probability $1$ (and, thus, $\ket{\varphi_N}$ is irreducibly $D$-partite entangled with probability $1$), $R_\sigma^{(n)}\left(\ket{\varphi_N}\right)$ has a high probability of being vanishingly small when $N$ becomes sufficiently large. Hence it becomes asymptotically hard to certify that $R_\sigma^{(n)}\left(\ket{\varphi_N}\right)>0$.
\begin{prop}
    Let $\sigma \in S_D$, $n \geq 2$, and let $(\ket{\varphi_N})_{N \in \mathbb{N}^*}$ denote a sequence of $D$-partite Haar-random states of local dimension $N$ (\ie $\ket{\varphi_N}$ is a uniformly distributed pure state in $\left(\mathbb{C}^N\right)^{\otimes D}$). Then, for any $\varepsilon > 0$, we have
    \begin{equation}
    \lim_{N \to +\infty} \mathbb{P}\left( R_{\sigma}^{(n)}(\ket{\varphi_N})\geq \varepsilon \right) = 0\,.
    \end{equation}
\end{prop}
\begin{proof}
    For any $n\geq 1$, $G_n(\tilde X_\sigma)$ is a \textit{joint realignment moment} (see Sec.~3.1.6 of~\cite{Carrozza:2026qcf}); in particular, it belongs to the family of \textit{compatible graphs}, that were shown to obey a \emph{large-$N$ factorization} criterion in \cite{Carrozza:2026cci}. This allows one to prove that the random variable $\Tr_{G_n(\tilde{X}_\sigma)}(\ket{\varphi_N})$ concentrates around its average in the $N \to +\infty$ limit (see Prop.~5.2 of \cite{Carrozza:2026cci}). Furthermore, equation \eqref{eq:power_law_bound} together with Prop.~6.16 of \cite{Carrozza:2026qcf}, imply that the invariant $\Tr_{G_n(\tilde{X}_\sigma)}$ obeys a power law (Def.~6.15 of \cite{Carrozza:2026qcf}) \ie there exist $K >0$ and $r>0$ such that: $\forall \ket\psi \in (\mathbb{C}^N)^{\otimes D}$, $\Tr_{G_n(\tilde{X}_\sigma)}(\ket\psi) \geq K N^{-r}$. This in turn implies that 
    \begin{equation}
        \left\langle \ln \left(\Tr_{G_n(\tilde{X}_\sigma)}(\ket{\varphi_N})\right) \right\rangle \underset{N\to +\infty}{\sim} \ln \left\langle \Tr_{G_n(\tilde{X}_\sigma)}(\ket{\varphi_N})\right\rangle\,,
    \end{equation}
    by Prop.~5.4 of \cite{Carrozza:2026cci}. The right-hand side of this equation can be evaluated explicitly and gives $2(1-D)n \ln N+o(1)$ (see \eg the tables from Sec.~4.3 of~\cite{Carrozza:2026qcf}). Since the leading-order contribution is linear in $n$, we obtain
    \begin{equation}
            \lim_{N \to +\infty}\left\langle R_{\sigma}^{(n)}(\ket{\varphi_N}) \right\rangle = 0\,,
    \end{equation}
    and, $R_{\sigma}^{(n)}(\ket{\varphi_N})$ being non-negative, Markov's inequality finally implies
    \begin{equation}
        \mathbb{P}\left( R_{\sigma}^{(n)}(\ket{\varphi_N})\geq \varepsilon \right) \leq \frac{\left\langle R_{\sigma}^{(n)}(\ket{\varphi_N}) \right\rangle}{\varepsilon} \underset{N \to +\infty}{\rightarrow} 0\,.
    \end{equation}
\end{proof}

\paragraph{Conclusion.} We have shown that there is an obstruction to the existence of subadditive $\LOCC$-monotones that vanish on all biseparable states, but detect the genuine multipartite entanglement of a complete $\GHZ$ state. On the contrary, no such limitation exists if one only requires $\LO$-monotonicity. Indeed, we have shown that the quantities $\{R_\sigma^{(n)}\}$ (resp.~$\{R^{(n)}\}$) define GMEs that are both (sub)additive and $\LO$-monotone. They take the form of a difference of two logarithms of invariant polynomials of degree $2(D-1)n$ (so of degree $8$ in the simplest non-trivial case, corresponding to $D=3$ and $n=2$), and fall in the general category of \emph{multipartite entanglement spreads} introduced in \cite{Carrozza:2026jth}. Many other similar examples of GMEs obeying conditions~\ref{it:LO}, \ref{it:sub} and \ref{it:bisep} can be constructed with our combinatorial approach: indeed, it is easy to generate other pairs of colored graphs obeying the key combinatorial constraints of \eqref{eq:comb_constraint}, \eg by inserting suitable decorations in the graphs $G_n(\tilde{X})$. However, any such example will depend on invariants of higher degree than the ones considered in the present letter. The examples we focused on in the present work arguably provide the simplest general class of GMEs obeying conditions~\ref{it:LO} to \ref{it:bisep} currently available in the literature. For quantum systems of low dimensions, we expect them to be efficiently measurable in the laboratory or on a quantum computer, following the approach of \cite{Szalay:2026yli} (a precise analysis being deferred to future work).

Looking ahead, it would be interesting to construct other families of GMEs by more systematically investigating the constraints imposed by conditions~\ref{it:LO} to \ref{it:bisep} on natural entropy  Ans\"{a}tze. For instance, starting from an arbitrary linear combination of logarithms of trace-invariants, which automatically implements condition~\ref{it:sub}, what are the restrictions imposed by conditions~\ref{it:LO} and \ref{it:bisep}? In particular, do the R\'{e}nyi multi-entropies of \cite{Iizuka:2025ioc, Iizuka:2025caq} define $\LO$-monotones? A more exhaustive understanding of this problem would be valuable, and is left for future work. 

\medskip

\section{Acknowledgments}
\begin{acknowledgments}
    This work was supported by the  ANR JCJC project ``RTFPQuEnt" (ANR-25-CE40-5465). The IMB receives support from the EIPHI Graduate School (contract ANR-17-EURE-0002). L.L. also acknowledges support  from the  ANR PRC project  ``TAGADA" (ANR-25-CE40-5672). No generative AI was used at any stage of this work.
\end{acknowledgments}

\appendix

\section{\label{A:LOCC} APPENDIX A: Proof of Lem.~\ref{lem:LOCC}} 

Let $\pmb{p}=(s_1, s_2, \ldots, s_N, 0, \ldots)$ be an entanglement spectrum of rank $N$ (hence $s_1 \geq s_2 \geq \cdots \geq s_N >0$). Consider a party $c \in B_1 \cap B_2$, representing, say, Alice's laboratory. Alice starts by preparing an ancillary state $\ket{S}_c$ in her laboratory, expressed in a suitable basis as:
\begin{equation}
    \ket{S}_c = \sum_{i=1}^N \sqrt{s_i} \ket{i} \ot \ket{i} \ot \ket{i}\,.
\end{equation}
We also recall the definition of the generalized Bell basis on $\mathrm{span}\{\ket{a}\otimes \ket{i}\}_{1 \leq a,i \leq N}$ whose elements, defined for any $1\leq m,n \leq N$, take the form
\begin{equation}
    \ket{\Phi_{m,n}} = \frac{1}{\sqrt{N}} \sum_{k = 1}^N \mathrm{e}^{2i \pi kn/N} \ket{k} \ot \ket{k+m} \,.
\end{equation}
Whenever $N < k+m \leq 2N$ in this equation, $k+m$ is identified with $k+m-N$. Using this basis, we have for the initial state $\ket{\psi_\mathrm{in}}\eqdef \ket{\GHZ}_{B_1 , N} \ot \ket{\GHZ}_{B_2 , N}$:
\begin{widetext}
    \begin{align}
       \ket{\psi_\mathrm{in}}  &\simLU  \ket{\GHZ}_{B_1 , N} \ot \ket{\GHZ}_{B_2 , N} \ot \ket{S}_c \,, \\
        &= \frac{1}{N} \sum_{i=1}^N \sum_{a=1}^{N} \sum_{b=1}^{N} \sqrt{s_i} \ket{a}^{\ot (\abs{B_1} - 1)} \ot \ket{b}^{\ot (\abs{B_2} - 1)} \ot \underbrace{\pa{\ket{a} \ot \ket{i}} \ot \pa{\ket{b} \ot \ket{i}} \ot \ket{i}}_{\mathrm{subsystem}\;c} \,, \\
        &= \frac{1}{N^2} \sum_{i = 1}^N \sum_{m_1,n_1 = 1}^N  \sum_{m_2,n_2 = 1}^N \sqrt{s_i}\,\theta_N^{(m_1-i)n_1 + (m_2-i)n_2}  \bigotimes_{v = 1}^2 \pac{(X^{-m_v}\ket{i})^{\ot (\abs{B_v} - 1)} \ot \underbrace{\ket{\Phi_{m_v,n_v}}}_{\mathrm{subsystem}\;c}} \ot \underbrace{\ket{i}}_{\mathrm{subsystem}\;c} \,,
    \end{align}
\end{widetext}
where $\theta_N \eqdef \exp(2i\pi/N)$, $X$ is the (unitary) Pauli operator defined by $X\ket{i} = \ket{i+1}$ for any $i\in\paa{1,\dots,N}$. We observe that the Bell basis elements appearing in the last equations are all local to Alice's laboratory. Thus, she may perform the $\LO$ operation consisting of the projective measurement given by the $N^4$ mutually orthogonal projectors acting on those subsystems:
\begin{equation}
    P_{\substack{m_1,n_1 \\ m_2,n_2}} \eqdef \ket{\Phi_{m_1,n_1}} \bra{\Phi_{m_1,n_1}} \ot \ket{\Phi_{m_2,n_2}} \bra{\Phi_{m_2,n_2}} \,.
\end{equation}
Alice can then share the outcome $(m_1,n_1,m_2,n_2)$ via classical communication with the other local laboratories. The resulting pure state at the end of the protocol is 
\begin{equation}
    \ket{\psi_\mathrm{out}} \simLU \ket{\sGHZ}_{B,\,\pmb{p}}\,,
\end{equation}
and an $\LU$-correction can thus be applied to obtain $\ket{\sGHZ}_{B,\,\pmb{p}}$ as a net result of this $\LOCC$ protocol. \hfill $\square$

\section{\label{B:IFF} APPENDIX B: Proof of Prop.~\ref{prop:iff_3qubits}} 

The implication ($\Leftarrow$) follows from Cor.~\ref{cor:if_GME}. For the direction ($\Rightarrow$), we prove the contrapositive. Let $\ket \psi \in \bb{C}^2 \ot \bb{C}^2 \ot \bb{C}^2$ such that $R_\sigma^{(n)}(\ket \psi) = 0$. We aim to prove that $\ket\psi$ is not irreducibly tripartite entangled which, in this system of three qubits, is equivalent to $\ket\psi$ being \emph{biseparable}.

Without loss of generality, we can assume that $\ket\psi$ takes the following canonical form \cite{Acin:2000jx,Acin:2000stw}:
\begin{align}
    \ket \psi =& \sqrt{\eta_0} \mathrm{e}^{i \phi} \ket{000} + \sqrt{\eta_1} \ket{001} + \sqrt{\eta_2} \ket{010} \nonumber \\
    &+ \sqrt{\eta_3} \ket{100} + \sqrt{\eta_4} \ket{111} \,, \label{eq:canonical_form}
\end{align}
where $0 \leq \eta_i \leq 1$, $0 \leq \phi < \pi$ and $\displaystyle \sum_{i=0}^4 \eta_i = 1$. Indeed, there is at least one state of the form \eqref{eq:canonical_form} in any $\LU$-orbit of a three-qubit system \cite{Acin:2000jx}. Likewise, we will assume, for definiteness, that $\sigma=(1)(2)(3)$.

Next, we remark that $R_\sigma^{(n)}(\ket \psi) = 0$ is equivalent to $A_\psi (A_\psi)^\dagger$ being rank-one, or (via the singular value decomposition) to $A_\psi$ itself being rank-one, where here $A_\psi \eqdef \Tr_{M_\sigma}(\ket\psi)$ with: 
\begin{equation}
    M_\sigma = \vcenter{\hbox{\includegraphics[height = 1.5cm]{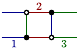}}}\,.
\end{equation}
Identifying the basis $\left( \ket{0}, \ket{1}\right)$ of $\mathbb{C}^2$ to the dual basis $\left( \bra{0}, \bra{1}\right)$ of $\left(\mathbb{C}^2\right)^*$, we can describe $$A_\psi: \mathbb{C}^4\simeq \mathbb{C}^2 \otimes \left(\mathbb{C}^2\right)^* \to \left(\mathbb{C}^2\right)^* \otimes \mathbb{C}^2\simeq \mathbb{C}^4$$ as a $4\times 4$ matrix in the basis $(\ket{00}, \ket{01}, \ket{10}, \ket{11})$. Since $A_\psi$ is rank-one, its $2 \times 2$ minors must all vanish. Explicitly evaluating some of those minors in terms of the canonical parameters from \eqref{eq:canonical_form}, we will find that $\ket\psi$ is necessarily biseparable.

To begin with, a computation shows that
\begin{equation}
    \left|\begin{array}{rr}
       \pa{A_\psi}_{33}  & \pa{A_\psi}_{34} \\
       \pa{A_\psi}_{43}  & \pa{A_\psi}_{44}
    \end{array}\right| = \sqrt{\eta_1 \eta_3} \pac{\eta_0 \eta_4^2 + (\eta_2 + \eta_4)(\eta_1 \eta_3 + \eta_4^2)}\,,
\end{equation}
and since this quantity vanishes, one must either have: (a)~$\eta_0 \eta_4^2 + (\eta_2 + \eta_4)(\eta_1 \eta_3 + \eta_4^2) = 0$, or (b)~$\eta_1 \eta_3=0$. Recalling that $\eta_i > 0$ for any $i$, condition (a) leads to $\eta_4 = 0$, as well as $\eta_j = 0$ for some $j\in \{1,2,3\}$. In that case, it is thus clear that $\ket\psi$ (as defined in \eqref{eq:canonical_form}) is biseparable.

If, on the other hand, condition (b) holds, one must either have $\eta_1 = 0$ or $\eta_3 = 0$. By symmetry of the canonical form, we can assume without loss of generality that $\eta_1 = 0$. In that case, one can compute another minor to show that:
\begin{equation}
    0= \left|\begin{array}{rr}
       \pa{A_\psi}_{32}  & \pa{A_\psi}_{34} \\
       \pa{A_\psi}_{42}  & \pa{A_\psi}_{44}
    \end{array}\right| = \sqrt{\eta_2}  \eta_3  \eta_4^{5/2}\,.
\end{equation}
We are thus led to one of the following two subcases: (b1) $\eta_1 = 0$ and $\eta_4 = 0$, or (b2) $\eta_1= 0$ and $\eta_j =0$ for some $j \in \{2,3\}$. In subcase (b1), the state $\ket\psi$ of \eqref{eq:canonical_form} is clearly biseparable. In subcase (b2), we can assume without loss of generality that $\eta_1 = \eta_2 = 0$. By means of an explicit evaluation of yet another minor, we then find that: 
\begin{equation}
0=     \left|\begin{array}{rr}
       \pa{A_\psi}_{11}  & \pa{A_\psi}_{14} \\
       \pa{A_\psi}_{41}  & \pa{A_\psi}_{44}
    \end{array}\right| = \eta_0 \eta_4^2(\eta_0 + \eta_3)\,.
\end{equation}
Hence, either $\eta_4 = 0$, or $\eta_0 = 0$. In either case, the state $\ket\psi$ is biseparable.\hfill $\square$

\bibliography{Biblio.bib}

\end{document}